# Assessing Expertise in Introductory Physics Using Categorization Task

Andrew Mason and Chandralekha Singh
Department of Physics and Astronomy, University of Pittsburgh, Pittsburgh, PA 15260

## Abstract

The ability to categorize problems based upon underlying principles, rather than surface features or contexts, is considered one of several proxy predictors of expertise in problem solving. With inspiration from the classic study by Chi, Feltovich, and Glaser [1], we assess the distribution of expertise amongst introductory physics students by asking three introductory physics classes, each with more than a hundred students, to categorize mechanics problems based upon similarity of solution. We compare their categorization with those of physics graduate students and faculty members. To evaluate the effect of problem context on students' ability to categorize, two sets of problems were developed for categorization. Some problems in one set included those available from the prior study by Chi et al. We find a large overlap between calculus-based introductory students and graduate students with regard to their categorizations that were assessed as "good". Our findings, which contrast from those of Chi et al., suggest that there is a wide distribution of expertise in mechanics among introductory and graduate students. Although the categorization task is conceptual, introductory students in the calculus-based course performed better than those in the algebra-based course. Qualitative trends in categorization of problems are similar between the non-Chi problems and problems available from the Chi study [1] used in our study although the Chi problems used are more difficult on average.

# I. Introduction

The nature of expertise and the transition from novice to expert is of interest to many researchers and practitioners. While many cognitive scientists and education researchers have focused on unraveling the nature of expertise, the community is still struggling with various facets of expertise [1-10]. These facets include identification of characteristics that are predictors of expertise, how expertise develops and whether this development is a gradual process or whether there are major boosts along the way in development as a result of certain types of exposure or scaffolding supports [11-20]. Physics has frequently been used as a domain in which the nature of expertise is investigated. This choice is partly because there is a well-defined hierarchical knowledge structure in physics, and because solving problems in physics involves applying a few fundamental laws which are expressed in precise compact mathematical forms in diverse situations.

## A. Background and Research on Expertise in Physics

An expert in physics is expected to have a functional understanding of physics [2-5]. He/she should make a connection between math and physics to interpret the physical significance of mathematical procedures and results, learn to convert a real physical situation into a mathematical model and apply mathematical procedures appropriately to solve physics problems beyond memorization of steps in a particular context, and estimate physical quantities and examine limiting cases in different situations as appropriate. Moreover, one cannot become an expert without developing productive attitudes about knowledge and learning in physics.

Reflection and sense-making is an integral component of expert behavior [2-6]. Experts monitor their own learning and are able to differentiate between what they know and what they do not know. They use problem solving as an opportunity for learning, extending, and organizing their knowledge. In order to become a physics expert, acquisition of content knowledge and development of a robust knowledge structure must go hand-in-hand with development of problem solving, reasoning and meta-cognitive skills.

Research suggests that the differences between expert and novice problem solving lie in both the level and complexity of how relevant knowledge is represented in memory (knowledge structure) and how heuristics are applied to solve problems (problem solving strategies) [21-30]. In general, experts in physics initially represent the problem space at a more abstract level (not very context dependent) and later focus on the specifics, while novices may immediately focus on the surface features or contexts of the problems. Experts start with visualizing the problem and performing the conceptual analysis and planning steps before resorting to the implementation of the plan. They employ representations (e.g., diagrammatic, tabular, graphical or algebraic) that make the problem solving task easier.

Novices, on the other hand may employ any of several problem solving approaches, which includes simply looking for plausible formulas without regard to applicability of concepts, or even none at all [31]. They also use a limited set of locally coherent resources when working together interactively [32]. A domain and task dependence on novice performance may also exist; for example, novice performance in categorization has been shown to depend on domain and task format for introductory electricity and magnetism problems in a different manner than mechanics problems [33].

Despite these general characteristics that emphasize experts and novices at two extremes of a continuum, expertise in physics can span a wide spectrum where a person with no knowledge of physics may be at one end of the spectrum and an adaptive expert who can apply his physics knowledge to solve novel complex problems may be at the other end [34-40]. Moreover, little is actually known about how expertise in physics develops as students make a transition from introductory to advanced courses, or whether the cognitive and meta-cognitive skills of physics graduate students are significantly superior to those of the physics and engineering majors in the introductory courses.

Research has shown that in order to become a world class expert in a domain, e.g., chess or music, one must deliberately practice for at least ten years [41]. However, introductory mechanics is often taught in a far smaller time frame, i.e. approximately one semester, and yet students may still be able to perform expert-like tasks over the course of a semester given that they study a very specific introductory domain within physics [42]. This expertise in a very limited domain is similar to that of a person who is learning a foreign language, who may become an expert in the alphabet and basic sentence structure in a short time. However, in order to master the entire literature in that language, the learner may require a significantly longer time. Introductory mechanics is not only a very limited part of the whole domain of physics but is also a very small part of the field of mechanics itself.

Physics graduate students have taken more advanced courses in mechanics which go beyond the topics covered in introductory physics. One pertinent question is related to how expertise develops and how beneficial these advanced mechanics courses are for developing a deeper understanding of introductory mechanics. Another issue is related to the fact that while introductory mechanics is relatively conceptual, advanced mechanics courses focus very heavily on mathematical tools. Conceptual understanding is almost never emphasized in these advanced physics courses under the assumption that graduate students should take the time to make the connections between conceptual and quantitative material themselves. There is no research on whether graduate students actually make such connections and deepen their conceptual knowledge structure of physics while learning mechanics in their graduate level courses, or whether a majority of graduate students never think about conceptual issues unless and until they become professors themselves. It is therefore useful to explore the extent to which the average performance of physics graduate students differs from that of introductory physics students if both groups are asked to perform conceptual tasks (e.g., categorization of problems) related to introductory mechanics content.

### B. Categorization and Expertise

Categorizing or grouping together various problems based upon similarity of solution is often considered a predictor of expertise [1,38,43]. An expert in physics may categorize many problems involving conservation of energy in one category and those involving conservation of linear momentum in another category, even if some of the problems involving both these conservation laws may have similar contexts and various problems involving conservation of energy may have different contexts. A good categorization based upon physics principles (deep features) may be challenging for beginning students because they may get distracted by the surface features or contexts of problems. Indeed, a significant body of research in psychology deals with concepts and categories [44-45].

In the classic study conducted by Chi, Feltovich and Glaser [1] (called the Chi study for convenience), eight introductory physics students in calculus-based courses were asked to categorize introductory mechanics problems based upon similarity of solution. They were also asked to explain the reasons for coming up with a particular category. Unlike experts who categorize problems based on the physical principles involved in solving them, introductory students, for example, categorized problems involving inclined planes in one category and pulleys in a separate category [1].

Analysis of data by Chi et al. [1] supported a theoretical framework that experts and novices categorize problems differently. It was found that the eight calculus-based introductory physics students (novices) were sensitive to the contexts or surface features of a physics problem and based their categorization on the problem's literal features. On the other hand, physics graduate students (experts) were able to identify physics principles applicable in a situation and categorize the problems based upon those principles. For example, 75%, 50%, 50%, and 38% of the novices had "springs", "inclined plane", "kinetic energy", and "pulleys" as one of the categories respectively. In addition, 25% of the experts used "springs" as a category, but "inclined plane", "kinetic energy", and "pulleys" were not chosen as category names by experts.

Other replication studies of the Chi study [1] expand upon findings about the expert-novice nature and how it pertains to problem solving. Veldhuis [36-37] employed a more detailed cluster analysis and a larger novice sample size to corroborate and extend the Chi study. The results indicate that categorization behavior of novices is more complex, and advanced novices' categorization tends to exhibit both deep structures and surface structures. Keith [46] found that novice students who explicitly make use of a general problem solving strategy will exhibit more expert-like categories. In Keith's study, however, the general problem solving strategy was part of instruction over the duration of an algebra-based physics course, and the number of participating students remained relatively low. It would be interesting to determine the categorization by students in large introductory physics classes and a large number of physics graduate students.

## II. Research Questions

With inspiration from the Chi study [1] on the categorization of introductory mechanics problems based upon similarity of solution, as well as the encouraging results of previous studies that built upon the Chi study, we compare the categorization of introductory physics problems in the calculus-based courses with introductory students in algebra-based courses, physics graduate students, and faculty members. Some of the data with a focus on the reasoning of the graduate students enrolled in a course for Teaching Assistants (TAs) when they categorized the problems from their own perspective and from the perspective of the introductory physics students they were teaching was discussed in an earlier paper [47]. The data presented here involves a larger number of introductory physics students including those in the algebra-based courses and includes two versions of the problem sets that students categorized (discussed later). Therefore, the data presented here can be used to answer a larger set of research questions including comparison with the Chi study which was not the focus earlier [47].

Within the theoretical framework that expert and novice categorizations differ, we chose to investigate a potentially wider spectrum in students' expertise in physics problem solving in large introductory physics classes that will not be captured by analyzing data from only 8 introductory student volunteers in the Chi study. We were motivated to assess the distribution of expertise in introductory physics by asking three introductory physics classes, each with more than a hundred students, to categorize mechanics problems based upon similarity of solution. The distribution of expertise in physics problem solving in these introductory physics classes is likely to represent a typical distribution in such courses.

### A. Comparison between Categorizations in Our Study and Chi study

Although no direct comparison is possible without access to all of the original problems in the Chi study, we may compare the distribution of students' expertise in categorizing physics problems in introductory physics classes with the eight introductory student volunteers in the Chi study. Specifically, we use two versions of the problem set, one of which involved the Chi problems available. We wish to investigate if there is a difference between the 8 Chi students and our student populations. Due to the small sample size in the Chi study, we refrain from using more detailed statistical analysis of Chi data on the grounds that the standard error will be too large to determine anything meaningful.

### B. Comparison between Chi and Non-Chi questions used in Our Study

To evaluate the effects of problem topic and of context within a mechanics topic on students' ability to categorize, two sets of problems were developed for categorization. Some problems in version II included all seven problems available from the Chi study. Only these problems were included because others were not available from Chi [48]. We may then use the version II problem set to specifically evaluate whether the Chi questions were easier or more difficult to categorize and whether there is a qualitative difference in the manner in which students categorized the Chi and non-Chi questions.

### C. Comparison of calculus-based introductory students with physics graduate students and faculty members

We also compare the distribution of calculus-based introductory physics students' expertise as manifested by their ability to categorize with those of physics graduate students and of faculty at the same university. Such comparisons are useful for assessing the extent to which the cohort of calculus-based introductory students is different from the graduate students (who have taken advanced mechanics courses) in the ability to categorize *introductory* mechanics problems. The comparison is likely to shed light on the extent to which categorization of problems is a predictor of expertise. The comparison between introductory students, graduate students and physics faculty members may also shed some light on whether the development of expertise as it pertains to the ability to categorize is gradual or whether it happens in spurts and there are major boosts from time to time, e.g., when one starts to teach. We note that we compare the calculus-based introductory physics students with the graduate students and faculty to keep our analysis similar to Chi study (in which 8 introductory physics students from calculus-based courses were involved). However, our next research question compares the students in the calculus-based courses with those in the algebra-based courses.

**D. Comparison of students in the calculus-based and algebra-based introductory physics classes**

We also hypothesized that students in calculus-based introductory courses will perform better than those in algebra-based introductory courses on the categorization task. Therefore we compare the respective performance of the introductory physics students in calculus-based and algebra-based courses. The calculus-based course is generally mainly taken by engineering majors, physics majors, and mathematics majors, while the algebra-based course is taken mainly by those with interest in health-related professions. The content of the calculus-based mechanics course is very similar (with the same topics covered in the same order) to that of the algebra-based mechanics course; the obvious difference is that the calculus-based courses use some calculus (although sparingly since most students in these physics courses are enrolled in the corresponding calculus course simultaneously). In general, the students in the calculus-based physics courses have a stronger mathematical background and display higher scores in the scientific reasoning skills test [49-54].

## III. Methodology

Below, we describe the procedure, materials and participants in the study.

### A. Procedure

All students and faculty members who performed the categorization task were provided the following instructions given at the beginning of the problem set:
* *Your task is to group the 25 problems below based upon similarity of solution into various groups on the sheet of paper provided. Problems that you consider to be similar should be placed in the same group. You can create as many groups as you wish. The grouping of problems should NOT be in terms of ``easy problems'', ``medium difficulty problems'' and ``difficult problems'' but rather it should be based upon the features and characteristics of the problems that make them similar. A problem can be placed in more than one group created by you. Please provide a brief explanation for why you placed a set of questions in a particular group. You need NOT solve any problems.*
* *Ignore the retarding effects of friction and air resistance unless otherwise stated.*

The sheet on which participants were asked to perform the categorization of problems had three columns. The first column asked them to use their own category name for each of their categories, the second column asked them for a description of each category that explains why problems within that category may be grouped together, and the third column asked them to list the problem numbers for the questions that should be placed in a category. Apart from these directions, neither students nor faculty were given any other hints about which category names they should choose.

### B. Necessary Differences in Procedure from Chi Study

The Chi study notes that students were asked to categorize the problems based upon similarity of solution. It should be noted that the exact written instructions to students were not given in the Chi paper and therefore are unfortunately lost [48]. While our instruction above also asks students to categorize the problems based upon similarity of solution, we had additional sentences in the instructions meant to clarify what they should do. As a preliminary check to make sure the problems were clear, we conducted individual interviews with a few introductory students and physics professors in which they were asked to categorize the problems using think-aloud protocol, and we found that all of them interpreted the instructions as intended (similar to the Chi study). Moreover, in the results section, we discuss that introductory students in general categorized problems better in our study than in the Chi study, which further supports the fact that our instruction is clear.

We also note that another difference between this study and the Chi study is that, since few students were involved in the Chi study, students were given each of the problems on index cards that could be sorted and placed in groups. In our study, which involved hundreds of students, categorization task was necessarily a paper-and-pencil task requiring students to write down their reasoning as well as their categories. Based upon the nature of the task, we do not anticipate that the performance of an individual with a certain level of expertise in mechanics (as manifested by categorization of problems) will be significantly affected by either of these implementation strategies.

Furthermore, in the Chi study, a record of how much time each student took to perform categorizations was maintained. In an in-class study with a large number of students, it was not practical to keep track of time. Instead, all students (introductory and graduate) had a full class period (50 minutes) to perform the categorization.

### C. Materials

Below, we describe the two versions of problem set used for categorization and the considerations in the selection of the problems and text.

#### 1. Two problem set versions

The Appendix includes all of the questions in the two versions of the problem sets given to the participants. Each version of the problem set contained 25 mechanics problems, 15 of which were included in both sets. The remaining 10 were unique for each problem set as discussed below.

The context of the 25 mechanics problems varied. Only 7 problems (called Chi problems for convenience) from the Chi study were known to us because they were the only ones mentioned in the Chi study and thus identifiable by the problem numbers from the third edition of introductory physics textbook by Halliday and Resnick (1974 edition). Personal communication with the lead author suggested that the problems in the original study not mentioned in their paper had been discarded and were not available [48]. In version I, which did not include any of the Chi problems, all of the 25 problems were mechanics problems developed by us (none were from the Chi study). However, the problems were on sub-topics similar to those chosen in the Chi study (rotational kinematics and dynamics was excluded in this version). The topics included one- and two-dimensional kinematics, dynamics, work-energy theorem, and impulse-momentum theorem and were distributed between these different topics as evenly as possible. Version II, which included the 7 Chi problems, also had 3 non-Chi questions on rotational kinematics and dynamics (beyond uniform circular motion) and angular momentum. The purpose of including additional (non-Chi) problems on rotational motion was to attempt to match these questions to the related Chi questions (e.g., #10 and #11 in Appendix version II) by deep structure, and thus eliminate the possibility that the Chi questions would stand out by being the only questions dealing with rotational kinematics and dynamics. Chi problems were included in version II in order to evaluate how students performed on those problems compared to the non-Chi problems. Version I had 10 problems that were different from the 7 Chi problems and the 3 rotational problems. Comparison of students' performance on the two versions was useful to evaluate which version was more challenging.

### 2. Considerations for problem choices and text

Many questions related to work-energy and impulse-momentum concepts were adapted from an earlier study [55] and many questions on kinematics and dynamics were chosen from other earlier studies [56-58] because the development of these questions and their wordings had gone through rigorous testing by students and faculty members. Some questions could be solved using one physics principle, e.g. conservation of mechanical energy, Newton's second law, or conservation of momentum. The first two columns of Table 1 show question numbers and examples of primary categories in which each question of the problem set version I (not involving the Chi problems) can be placed (based upon the physics principle used to solve each question). Questions 4, 5, 8, 24 and 25 are examples of problems that require the use of two principles to solve (see the Appendix). For example, Questions 4, 8, and 24 can be grouped together in one category because they require the use of conservation of mechanical energy and momentum. Similarly, the first two columns of Table 2 show question numbers and examples of the primary categories in which each of the 10 new problems of the problem set version II (involving the seven Chi problems) can be placed.

We note that the 7 available Chi questions used in this study involved non-equilibrium applications of Newton's laws, rotational motion or the use of two physics principles. While some of our problems also covered the same topics and had similar features, it is difficult to predict the exact match with other topics covered by other Chi problems (that are not available) even though they were also from the specific domain of mechanics. In choosing our own problems, we tried to cover the topics from the chapters in introductory mechanics and we also included problems with various levels of difficulty in solving them. For example, two part problems or non-equilibrium problems are more challenging than one part problems or equilibrium problems. Moreover, rotational motion is excluded from version I but is included in version II.

Moreover, as noted earlier, we often selected mechanics problems that had undergone rigorous testing by faculty members and students for unambiguous easy to interpret wording (although they were often adapted from the same textbook used in the Chi study) because students and faculty members sometimes find the wording of the textbook problems confusing. The problem context was a major consideration in the design and selection of problems. For example, there were several problems dealing with inclined planes in both versions. Also, version I had several problems with balls being shot or dropped off of cliffs whereas version 2 did not have these (such problems were replaced either by the Chi problems or additional rotational motion problems for comparison). Incidentally, some of the physics faculty members were given some of the Chi problems that were also used in our study and asked to categorize them while thinking aloud. Some of the faculty members pointed out that the wording of problem 14 (version II) could be made clearer if the man "started from rest" which was not mentioned. Also, the faculty members pointed out that problem 18 (version II) did not mention the coefficient of static friction, which was relevant for determining whether the block will come down from the highest point on the inclined plane where it is momentarily at rest. On the other hand, the criteria that guided our choice of items from introductory mechanics to include in the questionnaire were the same as those for the sorting task in the Chi study.

One difference between the Chi-problems and non-Chi problems used in our study is that some of the non-Chi problems included diagrams. These diagrams were included because students can misinterpret some verbal problems without diagrams which have a complicated arrangement of objects (see Appendix for examples of problems with diagrams). The inclusion of diagrams in those problems was supposed to make the situations presented in the problems easier to interpret so that students will not make errors in categorization due to incorrect interpretation of the problem situation. No diagram was included in the problems in which we did not intend any difficulty in interpreting the physical situation presented in the problems. Theoretically, one may hypothesize that those driven by the surface features of the problem may be adversely affected by the diagrams because the diagrams may draw their attention to such features, e.g., inclined plane. Contrary to this expectation, as discussed later in the result section, we found that introductory students in our study were significantly less likely to select inclined plane as a category than in the Chi study.

We also note that while it is very difficult to make problems with *identical* surface features and different deep features, there are very few fundamental laws in physics and problem solving involves applying those few principles in diverse situations. As can be seen from Tables 1 and 2, a majority of the problems we selected had very different contexts but they can be grouped in a few categories based upon the deep-structures (based upon a few laws of physics). We further note that, according to the Chi paper, in Study 2 "a set of 20 problems was constructed in which surface features were roughly crossed with applicable physical law". However, they note "Clearly, some problems could be solved using approaches based on either of two principles, force and energy, and in fact Judkis (an engineering student who was considered an expert in Chi study and consulted while selecting problems) solved them both ways. In these cases, the problem is listed under the principle he judged to yield the simplest or most elegant solution..." (page 131, Ref. [1]). We discussed the issue about whether elegance should be the criteria used for determining expert-novice behavior with several physics faculty members. Faculty members were not in agreement and many believed that as long as one categorizes a problem based upon how to solve it correctly, it is an "expert-like" categorization. We therefore did not pursue replicating "Study 2" from the Chi study.

C. **Participants**

Two algebra-based introductory physics classes (with 109 and 114 students) and one calculus-based introductory physics class (with 180 students) carried out the categorization task in their recitation classes. We note that all relevant concepts in the problem sets were taught in the introductory physics courses (whether it was algebra-based or calculus-based). All introductory students were told that they should try their best but they were given the same bonus points for doing the categorization task regardless of how expert-like their categorizations were. The 21 physics graduate students who carried out the categorization task were enrolled in a course for Teaching Assistants (TAs), and they performed the categorization in the last class of the semester. The seven physics faculty members who categorized the problems were asked to complete the task when convenient for them and return it to the researchers as soon as possible.

We note that one of the two algebra-based classes (with 109 students) was the class which was given the version II of the problem set (which included the Chi problems) to categorize. This version was used in order to evaluate whether there was any inherent difference in categorizing the Chi problems as opposed to the non-Chi problems and whether students' performance on the two problem sets was qualitatively similar. We also note that since the introductory students in the Chi study were 8 volunteers who responded to an advertisement, we are unsure whether they were enrolled in the introductory mechanics course in the same semester when they performed the categorization task or had taken introductory physics earlier.

In addition to the written categorization task administered to undergraduate and graduate students in various classes and seven physics faculty members, we also conducted individual interviews with four introductory physics students and a few graduate students and physics faculty members. The individual discussions were helpful in understanding their thought processes while they categorized the problems. Interviews will be briefly summarized here and will be explored in detail in a future publication.

## IV. Results

We first discuss how the categories were evaluated as good, moderate or poor and how they were classified before discussing the findings.

### A. Evaluation of Categories

Although we had our own assumptions about which categories created by individuals should be considered good or poor, we validated our assumptions with other experts. We randomly selected the categorizations performed by twenty calculus-based introductory physics students and gave them to three physics faculty who had taught calculus-based introductory physics recently (and who are known to not rush and be very thorough in any task they are assigned) and asked them to decide whether each of the categories created by individual students should be considered good, moderate, or poor. We asked them to mark each row which had a category name created by a student and a description of why it was the appropriate category for the questions that were placed in that category. If a faculty member rated a category created by an introductory student as good, we asked that he/she cross out the questions that did not belong to that category. The agreement between the ratings of different faculty members was better than *95%*.

We used faculty ratings as a guide to bin the categories created by everybody as good, moderate, or poor. Thus, a category was binned as "good" only if it was based on the underlying physics principles. We typically binned "conservation of energy" or "conservation of mechanical energy" as good categories. "Kinetic energy" is binned as a moderate category if students did not explain that the questions placed in that category can be solved using mechanical energy conservation or the work energy theorem. We binned a category such as "energy" as good if students explained the rationale for placing a problem in that category. If a secondary category such as "friction" or "tension" was the only category in which a problem was placed and the description of the category did not explain the primary physics principles involved, it was binned as a "moderate" category. Categories that were based upon surface features of the problems were binned as "poor." Examples of poor categories include "ramp" for objects on inclined surfaces, "pendulum" for objects tied to string, or "angular speed" if one must solve for angular speed (as opposed to a category based on principles such as rotational kinematics, rotational dynamics, angular momentum conservation). Table 1 shows examples of the primary and secondary categories and one commonly occurring poor/moderate category for each question given in the categorization task. We note that as can be seen from the instructions given, we did not ask for the primary and secondary categories explicitly but these two subgroups were determined based upon discussions with the faculty members.

  More than one principle or concept may be useful for solving a problem. The instructions specified that students could place a problem in more than one category. Because a given problem can be solved using more than one approach, categorizations based on different methods of solution that are appropriate were binned as "good" (e.g., see Table 1). For some questions, "conservation of mechanical energy" may be more efficient, but the questions can also be solved using one- or two-dimensional kinematics for constant acceleration.

  For questions that required the use of two major principles, those who categorized them in good categories either made a category which included both principles such as "conservation of mechanical energy" and "conservation of momentum" or placed such questions in two categories created by them -- one called "conservation of mechanical energy" and the other called "conservation of momentum". If such questions were placed only in one of the two categories, it was not binned as a good category; rather it was binned as a moderate category (this scoring scheme is not shown in Table 2 for clarity but was used in scoring individuals).

  We note that this way of scoring (good, moderate, and poor) can be compared to other categorization studies that claim to differentiate between deep and surface features in that all of the novice categories, e.g., from the Chi study (see Table 3 to be discussed later), would be classified as poor categories in the present study. However, while a majority of the expert categories in the Chi study would be classified as good categories in the present study, a few of them may fall in our moderate categories. In particular, if there were two fundamental principles required to solve a problem and the problem was placed in only one of those categories, we considered the categorization as moderate. Also, as discussed earlier, when the category names were vague, we determined whether it was good or moderate based upon the explanations provided. The Chi study does not clarify these issues although some of their expert category names cannot clearly be labeled or classified as based upon deep features (in Chi study, any category name that was mentioned by a graduate student was immediately taken to be based upon deep features).

### B. Classification of Categories

Classification of categories created by each individual consisted of placing each category by each person into a matrix which consisted of problem numbers along the columns and categories along the rows. In essence, a "1" was placed in a box if the problem appeared in the given category and a "0" was placed if the opposite was true. For example, for the 109 students in the algebra-based course who categorized version II of the problem set, an average of 7.02 categories per student was created. We recorded 82 proto-categories which were later reinterpreted into 59 categories. The latter process was carried out because many categories were interpreted to be paraphrases of other categories (e.g., ramp and inclined plane were taken to be the same categories).

We present Figures 1-4 for the categories that were binned as "good" by various student and faculty groups. We will discuss these figures later but we note that if a figure shows that 60% of the questions were placed in a good category by a particular group (calculus-based introductory students, algebra-based introductory students, graduate students, or faculty), it means that the other 40% of the questions were placed in the moderate or poor categories. An additional way to analyze the data would be to come up with an overall score for each participant (1 point for placing a problem in a good category, 0.5 if moderate, and 0 if poor) and then calculate an average score for each group. Such an analysis will be pursued in the future analysis of data.

### C. Comparison between Categorizations in Our Study and Chi study

Table 3 shows the list of categories that experts and novices created in the Chi study. The Table also includes the percentages of five different groups in our study who chose each of the categories created by experts and novices in the Chi study: both novices and experts in the Chi study, and then the introductory physics students in the calculus-based course and two algebra-based courses. The "cannot classify/omitted" category in Table 3 lists the percentage of students who noted they could not classify or skipped at least one question on the problem set. For version II of the test given to the algebra-based introductory physics class, which included seven Chi problems, we have two separate columns in Table 3 showing the categorization for only those seven questions and for all questions in version II.

Table 3 shows that the percentage of introductory students in our study who selected "ramps" or "pulleys" as categories (based mainly upon the surface features of the problem rather than the physics principle required to solve the problem) is significantly less than in the Chi study. One reason could be the difference in questions that were given to students in the two studies. In our study using version I of the problem set, introductory students sometimes categorized questions 3, 6, 8, 12, 15, 17, 18, 22, 24, and 25 as ramp problems, questions 6 and 21 as spring problems (question 21 was categorized as a spring problem by introductory students who associated the bouncing of the rubber ball with a spring-like behavior) and question 17 as a pulley problem. The lower number of introductory students referring to "springs" or "pulleys" as categories in our study could be due to the fact that there were fewer questions that involve springs and pulleys than in the Chi study. However, Table 3 shows that "ramp" was also a much less popular category for introductory students in our study (for version II, 19% chose this category for Chi problems and 24% for non-Chi problems) than in the Chi study, in which 50% of the students created this category and placed at least one problem into it. Similarly, Table 3 shows that kinetic energy was a novice category that was selected by 50% of the introductory students but in our study using both versions it was never more than 16%. Again, although we have 7 problems from the Chi study in version II, we cannot compare our data directly with theirs since most questions are different.

What is more surprising however is the fact that none of the 8 introductory physics students in the Chi study (see Table 3) chose Chi's expert categories, Newton's second law, energy principles, circular motion or linear kinematics as categories at all. On the other hand, Table 3 shows that in our study with version II, 18% selected Newton's second law for the 7 Chi-problems (22% for all), 31% selected energy principles for the 7 Chi problems (42% for all), 28% selected circular motion for the 7 Chi problems (29% for all) and 44% selected linear kinematics for the 7 Chi problems (51% for all). The fact that there were absolutely no introductory students choosing these categories in the Chi-study (see Table 3) but the percentage of students selecting these categories is quite large, *even for the Chi-problems used in our study*, is hard to reconcile even considering the small number of student volunteers in the Chi study. One factor contributing to this large difference may be that the student volunteers in the Chi study may not currently be taking the course (and may have forgotten the material), while the students in this study were concurrently enrolled in an introductory physics course. Further, we note that version II was only given to algebra-based introductory students who are generally worse at performing expert-like categorizations than the calculus-based introductory students (as discussed in the next section). The large discrepancies between the expert-like categorizations of problems in our study and the Chi study are likely to get even larger if we had given the 7 Chi-problems to the calculus-based group. One signature for this difference can be seen from Table 3 by comparing the last two columns which show that the algebra-based students in general produced less expert-like categorizations than the calculus-based students (calc-group) on version I of the problem set which did not include the Chi problems.

### D. Comparison between Chi and Non-Chi problems used in Our Study

Figure 1 shows a histogram of the percentage of questions placed in good categories by introductory students in the algebra-based course that used version II of the problem set that included the 7 available Chi problems. This figure compares the average performance on the categorization task when all problems were taken together vs. when Chi problems were separated out. We find qualitatively similar trends for the 7 Chi problems and non-Chi problems although the 7 Chi-problems were somewhat more difficult to categorize than the non-Chi problems (in each set of histograms for a given percentage of good category in Figure 1, the differences between the categorization of the Chi and non-Chi problems is within a standard deviation). We cannot infer anything further because we only had access to a few Chi problems (although they were all taken from the textbook).

Figure 2 is a histogram of the algebra-based introductory physics students (for both versions I and II of the problem set) who categorized various percentages of the 25 problems in "good" categories when asked to categorize them based on similarity of solution. Figure 2 also shows that version II involving Chi problems was categorized worse than version I. We note that in order to establish that there is no statistically significant difference between the two algebra-based physics classes, we performed t-tests between the two classes. The data for this was in the form of frequency of each problem being placed in a given category. For analysis we selected the 15 questions that both problem set versions had in common (see Table 2). First, the frequency of use for given categories was summed over all 15 problems and compared in a 1-way ANOVA test. The result was that there was no statistical significance between the two groups over all 15 problems that were present on both problem set versions ($p = 0.90$). In addition, t-tests were performed on each individual problem between the two populations to ensure that there were no individual problems that might suggest a difference between the two student populations. The p-value results ranged from 0.42 to 0.96, confirming that category frequency distributions were statistically similar for all 15 questions.

### E. Comparison of calculus-based introductory students with physics graduate students and faculty members

Figure 3 shows a histogram of the percentage of questions placed in good categories, and compares average performance on the categorization task of 21 graduate students and 7 physics faculty with the introductory calc-based group. Although the categorization of problems by the calc-based group is not on par with the categorization by physics graduate students, there is a large overlap between the two groups [47]. We note that in the Chi study the experts were graduate students and not physics professors, but Figure 3 suggests that there is a large difference between the graduate students and physics faculty in terms of their ability to categorize problems in good categories.

Overall, Figure 3 suggests that there is a wide distribution of performance amongst introductory students and graduate students in their ability to categorize mechanics problems and the definition of novice and expert used in the Chi study may not be appropriate, which is in keeping with the findings of Hardiman et al. [38]. In particular, the large overlap between graduate students (experts in the Chi study) and introductory physics students (novices in the Chi study) in Figure 3 appears to both corroborate and complement Keith's finding [46] about the mixture of expert-like and novice-like categorization among introductory students. In other words, not only is Keith's finding upheld about a smaller number of introductory students, but there is also a somewhat similar distribution in graduate student categorization.

### F. Comparison of students in calculus-based and algebra-based introductory physics classes

While the qualitative trends are similar for both groups, we find that categorizations by the introductory students in the calc-group are more expert-like than those by the students in the algebra-based course (algebra-group). In addition to the last two columns of Table 3 discussed earlier, the difference between the overall categorization by the calc-group compared to the algebra-group is evident in Figure 4, which is a histogram of the percentage of students in each group vs. the percentage of problems placed in good categories by each group for version I. The mean percentage of questions placed by the calc-group into good categories is 34.4% whereas the mean percentage by the algebra-group is 18.7%. This comparison between the calculus-based and algebra-based students suggests that the overlap between the algebra-based introductory students' and graduate students' categorization is likely to be less than that between the calculus-based introductory students' and graduate students' categorization.

### G. Are students unable to recognize relevant physics principles if their categories are not "good"?

In order to better understand the connection between categorization and expertise, we interviewed four introductory students and asked them to categorize problems during the interview while thinking aloud. We also asked a few graduate students and physics professors to categorize at least a subset of problems while thinking aloud in individual interview situations. While the interviews will be explored in more detail in a different publication, below we summarize some relevant findings.

All students interviewed could recall some concepts and create some categories that were based upon physics principles, but they also chose some categories that focused on the surface features of the problems. However, there is some evidence from the think-aloud interviews that moderate categories, e.g., "friction", were chosen as the categories although the students may have realized that a particular problem involving friction can be solved, e.g., using Newton's Second Law. Students sometimes deliberately chose "friction" as the major category instead of more expert-like categories based upon a fundamental physics principle because they felt the need to address specific details as opposed to the general physical principles. Upon asking for clarification, one introductory student who categorized a problem in "friction" category mentioned that he preferred "friction" category to the more general category of force or Newton's law because he found the term "force" to be vague and there were many problems that can be solved using Newton's law and the more specific description of different forces, e.g., friction, removed ambiguity. Some graduate student responses were also similar for similar situations.

Some physics professors were also specifically asked why they had placed a problem only in the "Newton's Second Law" category or "Work-Energy theorem" category as opposed to also including the specific forces or definition of work in the categorization. One professor responded that he thought that the task was about categorizing problems based upon the laws of physics and procedures for solving problems and, while the essential knowledge of forces and definitions of work were important to solve the problem, they were not the most fundamental issues that made the solution to the problems similar. In this sense, professors were more confident than students at any level that categorizing a problem in a very broad category based upon the physics principles would not make their categorization vague.

There is also evidence that some expert categories are not chosen or skipped by students because they may be viewed as superfluous in light of other categories that were already created by them (e.g., one student decided not to create a "kinematics" category after considering it because he already had an "energy" category in which he placed the problems that he would have also liked to place in the "kinematics" category). In comparison, physics professors were much more likely to place a problem in more than one category if it could be solved using two methods.

# V. Discussion

We find that the difference between the good categorizations performed by the physics professors and graduate students is much larger than the difference between graduate students and the calc-based group (see Figure 3). This finding contrasts with the Chi categorization study in which the introductory students and graduate students were found to be novices and experts, respectively.

We note that the physics professors pointed out multiple methods for solving a problem and specified multiple categories for a particular problem more often than graduate students and introductory students. Professors created secondary categories in which they placed problems that were more like the introductory students' and some graduate students' primary categories. For example, in version I of the problem set, in the questions involving tension in a rope or frictional force (see the Appendix) many faculty created these secondary categories called tension or friction, but also placed those questions in a primary category based on a fundamental principle of physics. For questions involving two major physics principles, for example, question 4 related to the ballistic pendulum, most faculty members categorized them in both "conservation of mechanical energy" and "conservation of momentum" categories in contrast to most introductory students in the calc-based group and many graduate students who either categorized it as an energy problem or as a momentum problem. The fact that most introductory students in the calc-based group and even many graduate students only focused on one of the principles involved to solve question 4 is consistent with an earlier study in which students either noted that this problem can be solved using conservation of mechanical energy or conservation of momentum but not both [55].

Many of the categories generated by the faculty, graduate students and introductory physics students were the same, but there was a difference in the fraction of questions that were placed in good categories by each group. What introductory students, especially those in the algebra-based courses, chose as their primary categories were often secondary categories created by the faculty members. Rarely were there secondary categories made by the faculty members, for example, a secondary category called "apparent weight", that were not created by students. There were some categories such as "ramps" and "pulleys" that were made by introductory physics students but not by physics faculty. Even if a problem did not explicitly ask for the "work done" by a force on an object, faculty members were more likely to create and place such questions which could be solved using the work-energy theorem or conservation of mechanical energy in categories related to these principles. This task was much more challenging for the introductory physics students who had learned these concepts recently (significantly more so for those in the algebra-based courses), and even for some graduate students. For example, it was easy to place question 3 in a category related to work because the question asked students to find the work done on an object, but placing problem 7 in the "work-energy" category was more difficult because students were asked to find the speed (see the Appendix).

Moreover, individual interviews with a few students in which they categorized problems while thinking aloud suggests that sometimes they categorized problems in concrete categories that were not considered good, e.g., friction, even though they knew that the problem can be solved using Newton's second law. Faculty members did not have such difficulty. Interviews suggest that due to their vast experience the faculty members had much more confidence in their categorizations based upon the fundamental laws of physics than students at the introductory or graduate level. Students at all levels sometimes second-guessed themselves and found categories based upon the laws of physics to be too general at times and preferred to use "friction" or "speed" as their categories rather than Newton's law or conservation of energy.

Based upon these findings, we believe that there is a connection between categorization and expertise, but it is unclear if it should be considered the hallmark of expertise. Further, we believe that rather than labeling people as novices and experts, it may be advantageous to think of them as located on a multi-dimensional continuum, with each dimension describing a different aspect of expertise. Ability to categorize problems can be considered one of those dimensions.

### A. Why isn't categorization by all faculty members "perfect"?

Although physics professors performed significantly better categorization than the graduate students, not all physics professors grouped all problems in good categories (see Figure 3). A closer look at the data suggests that faculty members' categorizations that were not good almost always had two types of errors:

(I) They inadvertently categorized a problem as being solvable using a particular principle of physics (e.g., work-energy theorem) when in fact another principle should be used to solve it (e.g., impulse-momentum theorem). We note that these types of errors have been reported previously when faculty members respond to conceptual questions [3, 59]. For example, Reif & Allen [3] asked introductory-level conceptual physics questions related to acceleration of a swinging pendulum to Berkeley physics professor and found that many of them answered the question incorrectly. In particular, they noted that the acceleration is zero when the pendulum bob is going through its mean position when in reality the acceleration is not zero. Such errors in answering conceptual questions is often due to the fact that faculty members are using their "compiled" knowledge about a class of problems (e.g., simple harmonic motion in the case of pendulum problem) to answer them rather than reasoning *explicitly* about the given situation. If instead of a pendulum, they were asked about a linear simple harmonic motion, e.g., a block attached to a spring, indeed the acceleration would be zero when going through its mean position. However, this result is not applicable to the pendulum since there is a centripetal acceleration. In fact, in one on one situation, when we asked some of the faculty members who had made errors in categorization to reconsider the categorizations of those problems or outline how they would solve them, they were able to correct their mistakes. Thus, if faculty members were asked to solve the problems explicitly rather than simply being asked to categorize them, they would most likely have realized that the principles they noted could be used to solve the problems were not appropriate in those situations. We make two additional related observations as follows:

a. In Study II in the Chi paper, one expert categorized a problem in a category different from the way Chi et al. had originally categorized them. They took the expert categorization as the good categorization (since there was only one expert who was given the task in their study it was not possible to verify the category with other experts) rather than their original categorization, assuming the expert could not go wrong. However, it is possible that the expert had made an error similar to our study.

b. According to the Study IV in the Chi paper and Hinsley et al. [60], "a problem can be categorized quickly (within 45 seconds, including reading time) and that it can often be tentatively categorized after reading just the first phrase of the problem.

According to this interpretation, a problem representation is not *fully* constructed until after the initial categorization has occurred. The categorization processes can be accomplished by a set of rules that specify problem features and the corresponding categories that they should cue." According to this interpretation, it is possible even for a physics professor to categorize an introductory physics problem incorrectly (if not done with great care) because the cues from the problem statements can sometimes be misleading and can bring out knowledge from memory that is not relevant for solving the problem, but such errors are likely to be detected when they actually solve the problems explicitly.

(II) Some faculty members categorized some problems that involved two physics principles in a category involving only one of those principles (such categorizations were not considered good). Similar to the point made earlier, such oversights are unlikely if they were asked to solve the problems explicitly.

### B. Why might the Calculus-based group perform better than the Algebra-based group on the Categorization Task?

A categorization task is primarily conceptual in nature and does not require quantitative manipulations. One may therefore wonder why students in the calc-based group performed more expert-like categorization than those in the algebra-based group. As noted in the introduction, the calculus-based course is predominantly taken by students who major in engineering, math and physics and are required to have reasonable mathematical and scientific reasoning skills, while the algebra-based introductory physics is mainly taken by the students with interests in careers in health professions who are majoring in biology, neuroscience, psychology and other disciplines and do not necessarily have strong mathematical skills or desire to learn concepts of engineering and physics.

As hypothesized earlier, one possible explanation for the difference between these two groups is based upon students' scientific reasoning abilities. Even conceptual reasoning of the kind needed for expert-like categorization in this study requires good scientific reasoning skills. Prior research has shown that the students in the calculus-based courses are better at conceptual reasoning and may be better at scientific reasoning skills [49-54]. The better mathematical preparation and scientific reasoning skills of the calculus-based students may reduce the cognitive load while learning physics and these students may not expend as much of their cognitive resources on processing information that is peripheral to physics itself, and may therefore have more opportunity to build a robust knowledge structure. If that is the case, students in the calculus-based classes may be able to perform better on conceptual tasks such as categorization than those in the algebra-based courses whose physics knowledge structure may not be as robust and skills in scientific reasoning about physical phenomena not as developed as the calc-based group. Another reason for the difference between the groups may be due to the fact that the calc-based group is more likely to have taken a physics class in high school and may have solved more mechanics problems than the students in the algebra-based group.

## VI. Conclusion

We were inspired by the classic categorization study by Chi et al. to investigate the distribution in introductory physics and graduate students' ability to categorize introductory mechanics problems. The study was conducted three decades after the Chi study in the classroom environment with several hundred introductory physics students. We asked individuals to categorize 25 mechanics problems based upon similarity of solution. Two versions of the problem sets were used in the study to investigate the impact of specific contexts of questions on categorization, with one version including the seven problems that were available from the Chi study.

      We find a large overlap between the categorizations performed by the calc-based group and the physics graduate students that were considered good. This large overlap in the performance of the two groups suggests that there is a wide distribution of expertise as assessed by the categorization task in both groups. Hence, it is not appropriate to classify all introductory physics students in calc-based courses as "novices" and all physics graduate students as "experts". We find that the categorization performed by physics faculty members was significantly better than that performed by the graduate students. Grouping all introductory physics students at the same level of expertise, or calling all graduate students "experts" misses a lot (if not most) of the features and essence of expertise.

      The overall qualitative trends in categorization were not strongly dependent on the version of the problem set given to the students (one of them involved the 7 Chi problems). While it is not possible to compare our data directly with that in the Chi study (most of their questions are no longer available), the percentage of introductory physics students who chose "surface-feature" categories such as "ramp" and "pulley" was significantly lower than the percentages reported in the Chi study. Even more striking is the fact that while none of the introductory students in the Chi study selected "expert" categories such as "Newton's second law", or "linear kinematics" etc. a significant number of introductory students did choose these categories in our study. Even if we restrict our study to the 7 problems common with Chi, the number of introductory students who selected such "expert" categories is significantly larger than zero in the Chi study (see Table 3). One issue that cannot be resolved here is the difference between investigations conducted in a classroom (ours) vs. that conducted outside the classroom with a few student volunteers (Chi study). The distribution of students' expertise in an in-class study is likely to reflect the distribution in a typical classroom (including high achieving and low achieving students). On the other hand, the distribution in an out of class study is more unpredictable and depends on the volunteer pool including issues such as how long ago they took the physics course.

      Our finding suggests that while expertise plays a role in categorization, and is a predictor of expertise, it is not appropriate to call all introductory students novices and all graduate students experts as in Chi study. In the future, it will be useful to investigate how categorization performance will differ if students are given the names of categories they could choose from (which would include both poor categories such as ramps and pulleys and good categories based upon the laws of physics) but were told that they need not use all of them and could even come up with their own categories. Future investigation might also explore similarities and differences in introductory students' and graduate students' responses if they were asked to solve the problems or at least outline a solution procedure rather than only being asked to do categorization. The earlier section describing individual interviews with a few students already shows hints that even those who do not categorize a problem in good categories may actually know what principles of physics is relevant for that problem. The opposite may also be true, in that those who categorize a problem correctly may have difficulty in delineating correct procedure for it.

# Acknowledgments

We thank Fred Reif and Bob Glaser for helpful discussions and thank the faculty members who administered the categorization task in their classes for their help.

# Appendix 1: Version I of the Problem Set

EPAPS: Appendix : Categorization Task: Version I

---
**Instructions**

- Your task is to group the 25 problems below based upon similarity of solution into various groups on the sheet of paper provided. Problems that you consider to be similar should be placed in the same group. You can create as many groups as you wish. The grouping of problems should NOT be in terms of "easy problems", "medium difficulty problems" and "difficult problems" but rather it should be based upon the features and characteristics of the problems that make them similar. A problem can be placed in more than one group created by you. Please provide a brief explanation for why you placed a set of questions in a particular group. You need NOT solve any problems.
- Ignore the retarding effects of friction and air resistance unless otherwise stated.

---

1. Harry Potter and Voldemort are wrestling inside a cart traveling east at a speed of 45 m/s directly toward an abyss. Harry then notices the danger and jumps backward due west off the cart. Ron who stands on safe ground in the back notices that Harry's velocity due west at the jump is 15 m/s relative to the ground. What is the speed of the cart after Harry jumps off it? The mass of the cart is 200 kg, Harry's mass is 60 kg, and Voldemort's mass is 80 kg.

2. Two identical stones, A and B, are shot from a cliff from the same height $h$ and with identical initial speeds $v_0$. Stone A is shot vertically up, and stone B is shot vertically down (see Figure). Which stone has a larger speed right before it hits the ground?

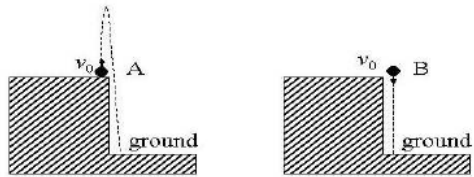

3. You want to lift a heavy block through a height $h$ by attaching a string of negligible mass to it and pulling so that it moves at a constant speed $v$. You have the choice of lifting it either by pulling the string vertically upward or along a frictionless inclined plane (see Figure). How much is the work done by the gravitational force in the two cases?

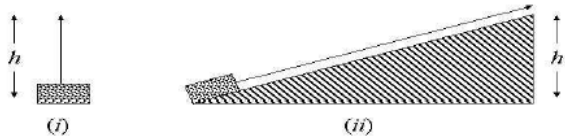

4. Two small spheres of putty, A and B, of equal mass, hang from the ceiling on massless strings of equal length. Sphere A is raised to a height $h_0$ as shown below and released. It collides with sphere B (which is initially at rest); they stick and swing together to a maximum height $h_f$. Find the height $h_f$ in terms of $h_0$.

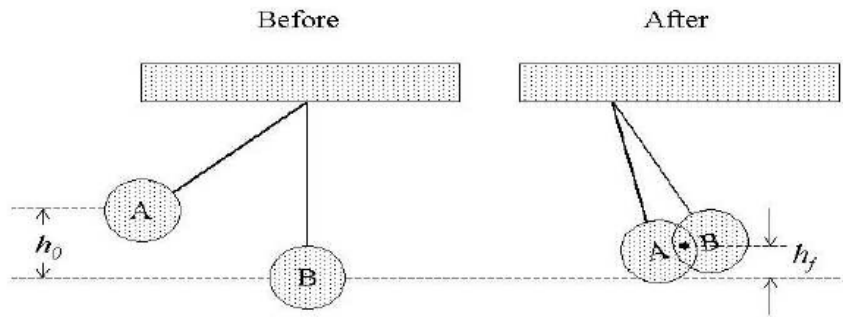

5. A family decides to create a tire swing in their backyard for their son Ryan. They tie a nylon rope to a branch that is located 16 m above the earth, and adjust it so that the tire swings 1 meter above the ground. To make the ride more exciting, they construct a launch point that is 13 m above the ground, so that they don't have to push Ryan all the time. You are their neighbor, and you are concerned that the ride might not be safe, so you calculate the maximum tension in the rope to see if it will hold. Calculate the maximum tension in the rope, assuming that Ryan (mass 30 kg) starts from rest from his launch pad. Is it greater than the maximum rated value of 2500 N?

6. In the figure below, a horizontal spring with spring constant $k_1 = 8$ N/m is compressed 20 cm from its equilibrium position by a 4 kg block. Then, the block is released. What would be the maximum compression of a spring ($k_2 = 5$ N/m) on the inclined plane when the 4 kg block presses against it? Assume that the track is frictionless and the distance from A to B is 50 cm where B is the edge of the uncompressed spring on the inclined plane.

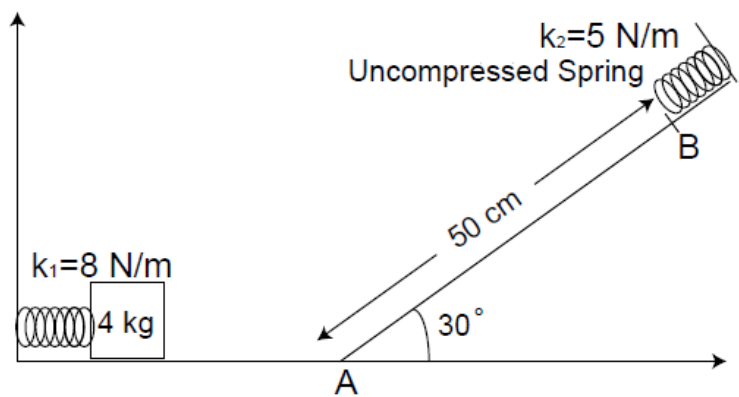

7. Two blocks are initially at rest on a frictionless horizontal surface. The mass $m_A$ of block A is <u>less</u> than the mass $m_B$ of block B. You apply the <u>same constant force $F$</u> and pull the blocks through the <u>same distance $d$</u> along a straight line as shown below (force $F$ is applied for the entire distance $d$). Compare the speed of the blocks after you pull them the <u>same distance $d$</u>.

$m_A < m_B$   TOP VIEW

A  $F \rightarrow$

B  $F \rightarrow$

START                FINISH

$\longleftarrow d \longrightarrow$

8. Your friend Dan, who is in a ski resort, competes with his twin brother Sam on who can glide higher with the snowboard. Sam, whose mass is 60 kg, puts his 15 kg snowboard on a level section of the track, 5 meters from a slope (inclined plane). Then, Sam takes a running start and jumps onto the stationary snowboard. Sam and the snowboard glide together till they come to rest at a height of 1.8 m above the starting level. What is the minimum speed at which Dan should run to glide higher than his brother to win the competition? Dan has the same weight as Sam and his snowboard weighs the same as Sam's snowboard.

9. Two identical stones, A and B, are shot from a cliff from the same height $h$ and with identical initial speeds $v_0$. Stone A is shot at an angle of $30^0$ above the horizontal and stone B is shot at an angle of $30^0$ below the horizontal. Which stone takes a longer time to hit the ground?

10. At amusement parks, there is a popular ride in which the floor of a rotating cylindrical room falls away, leaving the backs of the riders "plastered" against the wall. Suppose the radius of the room is 3.3 m and the speed of the wall is 10 m/s when the floor falls away. What is the minimum coefficient of friction that must exist between a rider's back and the wall, if the rider is to remain in place when the floor drops away?

11. Rain starts falling vertically down into a cart (of mass $M$) with frictionless wheels which is <u>initially</u> moving at a constant speed $V$ on a horizontal surface. The rain drops fall on the car with a speed $v$ and come to rest with respect to the cart after striking it. Find the speed of the cart when $m$ grams of rain water accumulate in the cart.

rain

frictionless wheels

12. In the track shown below, section AB is a quadrant of a circle of 1 m radius. A block is released at A and slides without friction until it reaches point B. The horizontal part is not smooth. If the block comes to rest 3 m from B, what is the coefficient of kinetic friction?

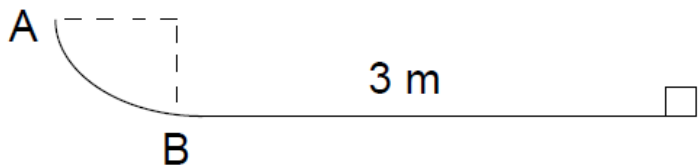

13. Three blocks ($m_1$=1 kg, $m_2$=2 kg, $m_3$=3kg) are in a straight line in contact with each other on a frictionless horizontal table (block with mass $m_2$ is in the middle). A constant horizontal force $F_H$=3 N is applied to the block with mass $m_1$. Find the forces exerted on $m_1$ by $m_2$ and on $m_2$ by $m_3$.

14. A ball is thrown from the top of a 35 m high building with an initial speed of 80 m/s at an angle of $25^0$ above the horizontal. Find the time it takes to reach the ground.

15. A cyclist approaches the bottom of a gradual hill at a speed of 15 m/s. The hill is 5 m high, and the cyclist estimates that she is going fast enough to coast up and over it without peddling. Ignoring friction and air resistance, find the speed at which the cyclist crests the hill? Neglect the kinetic energy of the rotating wheels.

16. A slingshot fires a pebble from the top of a building at a speed of 10 m/s. The building is 20 m tall. Ignoring air resistance, find the speed with which the pebble strikes the ground when the pebble is fired (I) horizontally, (II) vertically straight up.

17. The figure below shows two blocks on a frictionless inclined plane with an angle of inclination $\theta = 40^0$ and the two connected to each other via a massless rope. The rope that connects the two blocks goes around a frictionless massless pulley and is connected to a third block as shown. Find the magnitude of the tension force in the rope between blocks with mass $M_1$ and $M_2$ and the acceleration of the blocks.

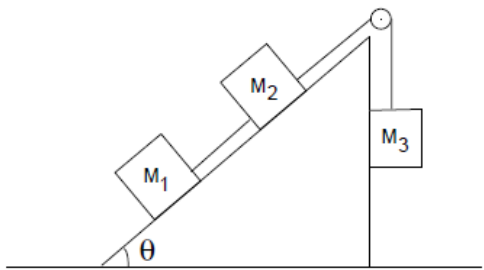

18. Two frictionless inclined planes have the same height but have different angles of inclinations of $45^0$ and $60^0$ with respect to the horizontal. You slide down from the top which is at a height $h$ above the ground on each inclined planes starting from rest. Find your speed at the bottom of the inclined planes in the two cases.

19. The brakes of your bicycle have failed, and you must choose between slamming into either a haystack or a concrete wall. Explain why hitting a haystack is a wiser choice than hitting a concrete wall.

20. Three balls are launched from the same horizontal level with identical speeds $v_0$ as shown below. Ball (1) is launched vertically upward, ball (2) at an angle of $60^0$, and ball (3) at an angle of $45^0$. In order of decreasing speed (fastest first), rank the speed each one attains when it reaches the level of the dashed horizontal line. All three balls have sufficient speed to reach the dashed line.

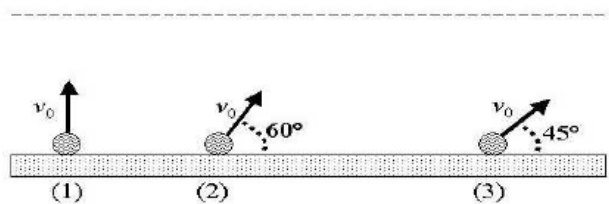

21. You drop two balls of equal mass, made of rubber and putty, from the same height $h$ above a horizontal surface (see Figure). The rubber ball bounces up after it strikes the surface while the putty ball comes to rest after striking it. Assume that in both cases the velocity of the ball takes the same time $\Delta t$ to change from its initial to its final value due to contact with the surface. During time $\Delta t$, which of the average forces $\overline{F}_R$ or $\overline{F}_P$ exerted on the surface by the rubber and putty balls, respectively, is greater?

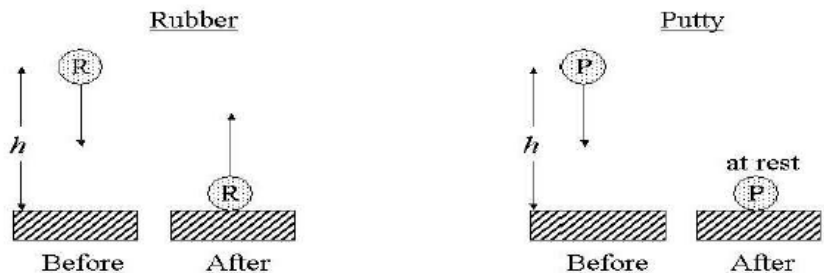

22. Two frictionless inclined planes have the same height but have different angles of inclinations of $45^0$ and $60^0$ with respect to the horizontal. You slide down from the top which is at a height $h$ above the ground on each inclined planes starting from rest. Find the time taken to reach the bottom in the two cases.

23. Two blocks are initially at rest on a frictionless horizontal surface. The mass $m_A$ of block A is less than the mass $m_B$ of block B. You apply the same constant force $F$ and pull the blocks through the same distance $d$ along a straight line as shown below (force $F$ is applied for the entire distance $d$). Rank the time taken to pull the two blocks by the same distance $d$.

$m_A < m_B$    TOP VIEW

A    F →

B    F →

|← d →|

START                    FINISH

24. You are standing at the top of an incline with your skateboard. After you skate down the incline, you decide to "abort", kicking the skateboard out in front of you such that you remain stationary afterwards. How fast is the skateboard travelling with respect to the ground after you have kicked it? Assume that your mass is 60 kg, the mass of the skateboard is 10 kg, and the height of the incline is 10 cm.

25. A friend told a girl that he had heard that if you sit on a scale while riding a roller coaster, the dial on the scale changes all the time. The girl decides to check the story and takes a bathroom scale to the amusement park. There she receives an illustration (see below), depicting the riding track of a roller coaster car along with information on the track (the illustration scale is not accurate). The operator of the ride informs her that the rail track is smooth, the mass of the car is 120 kg, and that the car sets in motion from a rest position at the height of 15 m. He adds that point B is at 5m height and that close to point B the track is part of a circle with a radius of 30 m. Before leaving the house, the girl stepped on the scale which indicated 55 kg (the scale is designed to be used on earth and displays the mass of the object placed on it). In the rollercoaster car the girl sits on the scale. According to your calculation, what will the scale show at point B?

# Appendix 2: Version II of the Problem Set

Categorization Task: Version II

> **Instructions**
>
> - Your task is to group the 25 problems below based upon similarity of solution into various groups on the sheet of paper provided. Problems that you consider to be similar should be placed in the same group. You can create as many groups as you wish. The grouping of problems should NOT be in terms of "easy problems", "medium difficulty problems" and "difficult problems" but rather it should be based upon the features and characteristics of the problems that make them similar. A problem can be placed in more than one group created by you. Please provide a brief explanation for why you placed a set of questions in a particular group. You need NOT solve any problems.
> - Ignore the retarding effects of friction and air resistance unless otherwise stated.

1. You drop two balls of equal mass, made of rubber and putty, from the same height $h$ above a horizontal surface (see Figure). The rubber ball bounces up after it strikes the surface while the putty ball comes to rest after striking it. Assume that in both cases the velocity of the ball takes the same time $\Delta t$ to change from its initial to its final value due to contact with the surface. Compare the average forces $\overline{F}_R$ and $\overline{F}_P$ exerted on the surface by the rubber and putty balls, respectively, during time $\Delta t$.

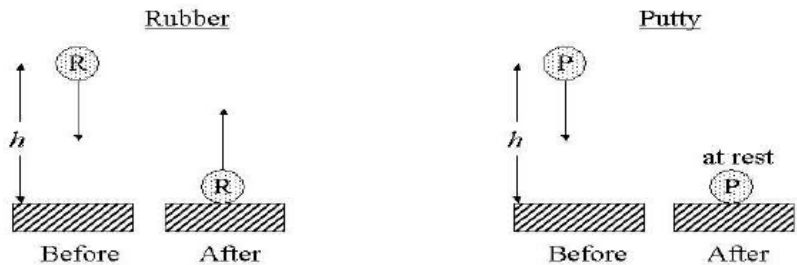

2. An ice skater is spinning on essentially frictionless ice with her arms extended. Then, she pulls her arms in close to her body, cutting her moment of inertia in half. There are no net external forces or torques on her. How will her angular speed change when she puts her arms in?

3. Your friend Dan, who is in a ski resort, competes with his twin brother Sam on who can glide higher with the snowboard. Sam, whose mass is 60 kg, puts his 15 kg snowboard on a level section of the track, 5 meters from a slope (inclined plane). Then, Sam takes a running start and jumps onto the stationary snowboard. Sam and the snowboard glide together till they come to rest at a height of 1.8 m above the starting level. What is the minimum speed at which Dan should run to glide higher than his brother to win the competition? Dan has the same weight as Sam and his snowboard weighs the same as Sam's snowboard.

4. Three blocks ($m_1$=1 kg, $m_2$=2 kg, $m_3$=3kg) are in a straight line in contact with each other on a frictionless horizontal table (block with mass $m_2$ is in the middle). A constant horizontal force $F_H$=3 N is applied to block with mass $m_1$. Find the forces exerted on $m_1$ by $m_2$ and on $m_2$ by $m_3$.

5. A ball is thrown from the top of a 35 m high building with an initial speed of 80 m/s at an angle of $25^0$ above the horizontal. Find the time it takes to reach the ground.

6. A cylist approaches the bottom of a gradual hill at a speed of 15 m/s. The hill is 5 m high, and the cyclist estimates that she is going fast enough to coast up and over it without peddling. Ignoring friction and air resistance, find the speed at which the cyclist crests the hill.

7. The figure below shows two blocks on a frictionless inclined plane with an angle of inclination $\theta = 40^0$ and the two connected to each other via a massless rope. The rope that connects the two blocks goes around a frictionless massless pulley and is connected to a third block as shown. Find the magnitude of the tension force in the rope between blocks with mass $M_1$ and $M_2$ and the acceleration of the blocks.

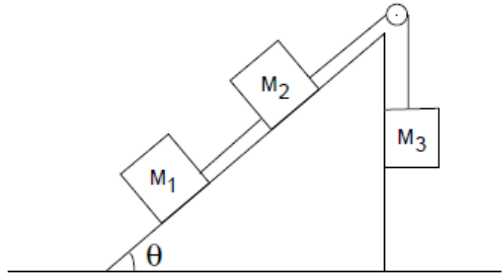

8. The brakes of your bicycle have failed, and you must choose between slamming into either a haystack or a concrete wall. Explain why hitting a haystack is a wiser choice than hitting a concrete wall.

9. You are standing at the top of an incline with your skateboard. After you skate down the incline, you decide to "abort", kicking the skateboard out in front of you such that you remain stationary afterwards. How fast is the skateboard travelling with respect to the ground after you have kicked it? Assume that your mass is 60 kg, the mass of the skateboard is 10 kg, and the height of the incline is 10 cm.

10. A heavy flywheel rotating on its axis is slowing down because of friction in its bearings. At the end of the first minute its angular speed is 0.90 of its angular speed $\omega_0$ at the start. Assuming constant frictional forces, find its angular speed at the end of the second minute.

11. A girl (mass M) stands on the edge of a frictionless merry-go-round (mass 10 M, radius R, rotational inertia I) that is not moving. She throws a rock (mass m) in a horizontal direction that is tangent to the outer rim of the merry-go-round. The speed of the rock, relative to the ground, is $v$. What is the speed of the girl after she throws the rock?

12. Two frictionless inclined planes have the same height but have different angles of inclinations of $45^0$ and $60^0$ with respect to the horizontal. You slide down from the top which is at a height $h$ above the ground on each inclined planes starting from rest. Find your speed at the bottom of the inclined planes in the two cases.

13. In the track shown below, section AB is a quadrant of a circle of 1 m radius. A block is released at A and slides without friction until it reaches point B. The horizontal part is not smooth. If the block comes to rest 3 m from B, what is the coefficient of kinetic friction?

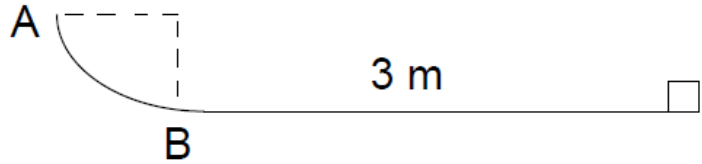

14. A man of mass $M_1$ lowers himself to the ground from a height $X$ by holding onto a rope passed over a massless frictionless pulley and attached to another block of mass $M_2$ on the other side. The mass of the man is greater than the mass of the block. With what speed does the man hit the ground after falling through a distance $X$?

15. A man of mass $M_1$ lowers himself to the ground from a height $X$ by holding onto a rope passed over a massless frictionless pulley and attached to another block of mass $M_2$ on the other side. The mass of the man is greater than the mass of the block. What is the tension in the rope?

16. A 2-kg block is forced against a horizontal spring of negligible mass, compressing the spring by 15cm. When the block is released, it moves 60 cm across a horizontal tabletop before coming to rest. The force constant of the spring is 200 N/m. What is the coefficient of friction between the block and the table?

17. A 4-kg block starts up a $30^0$ inclined plane with a kinetic energy of 128 J. How far will it slide up the plane if the coefficient of kinetic friction is 0.3?

18. A 2 kg block is given an initial speed of 4 m/s up an inclined plane starting from a point 2m from the bottom as measured along the plane. If the plane makes an angle of $30^0$ with the horizontal and the coefficient of friction is 0.2, with what speed will it reach the bottom of the plane?

19. After the fusion reaction is over, a star collapses under its own gravitational force into a neutron star, shrinking to 1/100th of its initial radius. Assume that there are no external torques on the star and that it loses no mass as it collapses. You may treat it as a uniform sphere. If its initial angular momentum and angular speed are $L_0$ and $\omega_0$, respectively, what are those variables after the collapse?

20. Two identical stones, A and B, are shot from a cliff from the same height $h$ and with identical initial speeds $v_0$. Stone A is shot vertically up, and stone B is shot vertically down (see Figure). Which stone has a larger speed right before it hits the ground?

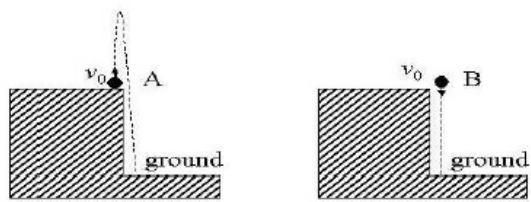

21. You dive from a diving board into the swimming pool. You want to make several somersaults before landing in the water. Should you tuck your body in or spread it out as you fall from the diving board? Explain your reasoning.

22. Two small spheres of putty, A and B, of equal mass, hang from the ceiling on massless strings of equal length. Sphere A is raised to a height $h_0$ as shown below and released. It collides with sphere B (which is initially at rest); they stick and swing together to a maximum height $h_f$. Find the height $h_f$ in terms of $h_0$.

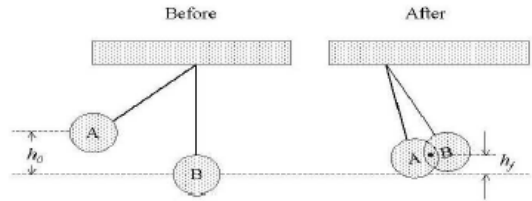

23. A family decides to create a tire swing in their back yeard for their son Ryan. They tie a nylon rope to a branch that is located 16 m above the earth, and adjust it so that the tire swings 1 meter above the ground. To make the ride more exciting, they construct a launch point that is 13 m above the ground, so that they don't have to push Ryan all the time. You are their neighbor, and you are concerned that the ride might not be safe, so you calculate the maximum tension in the rope to see if it will hold. Calculate the maximum tension in the rope, assuming that Ryan (mass 30 kg) starts from rest from his launch pad. Is it greater than the rated value of 2500 N?

24. At amusement parks, there is a popular ride in which the floor of a rotating cylindrical room falls away, leaving the backs of the riders "plastered" against the wall. Suppose the radius of the room is 3.3 m and the speed of the wall is 10 m/s when the floor falls away. What is the minimum coefficient of friction that must exist between a rider's back and the wall, if the rider is to remain in place when the floor drops away?

25. A friend told a girl that he had heard that if you sit on a scale while riding a roller coaster, the dial on the scale changes all the time. The girl decides to check the story and takes a bathroom scale to the amusement park. There she receives an illustration (see below), depicting the riding track of a roller coaster car along with information on the track (the illustration scale is not accurate). The operator of the ride informs her that the rail track is smooth, the mass of the car is 120 kg, and that the car sets in motion from a rest position at the height of 15 m. He adds that point B is at 5m height and that close to point B the track is part of a circle with a radius of 30 m. Before leaving the house, the girl stepped on the scale which indicated 55 kg (the scale is designed to be used on earth and displays the mass of the object placed on it). In the rollercoaster car the girl sits on the scale. According to your calculation, what will the scale show at point B?

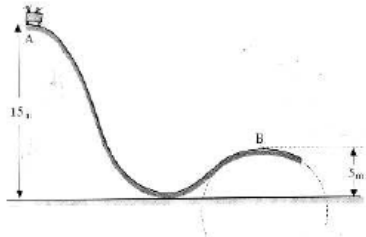

# Figures

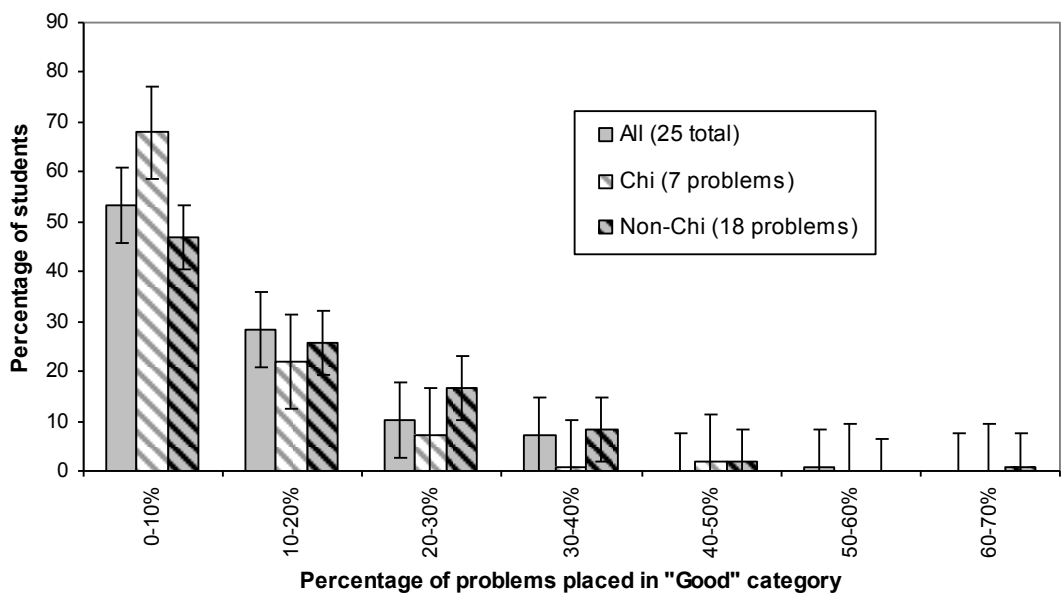

*Figure 1.* Histogram of algebra-based introductory physics students (total 109 students) who categorized various percentages of the 25 problems in version II of the problem set in "good" categories when asked to categorize them based on similarity of solution. The 7 Chi problems were categorized worse than the other 18 problems showing that the nature of introductory physics questions is important in students' ability to categorize them. The percentages of students for all 25 problems taken together are also shown. The error bars in all graphs show the standard error.

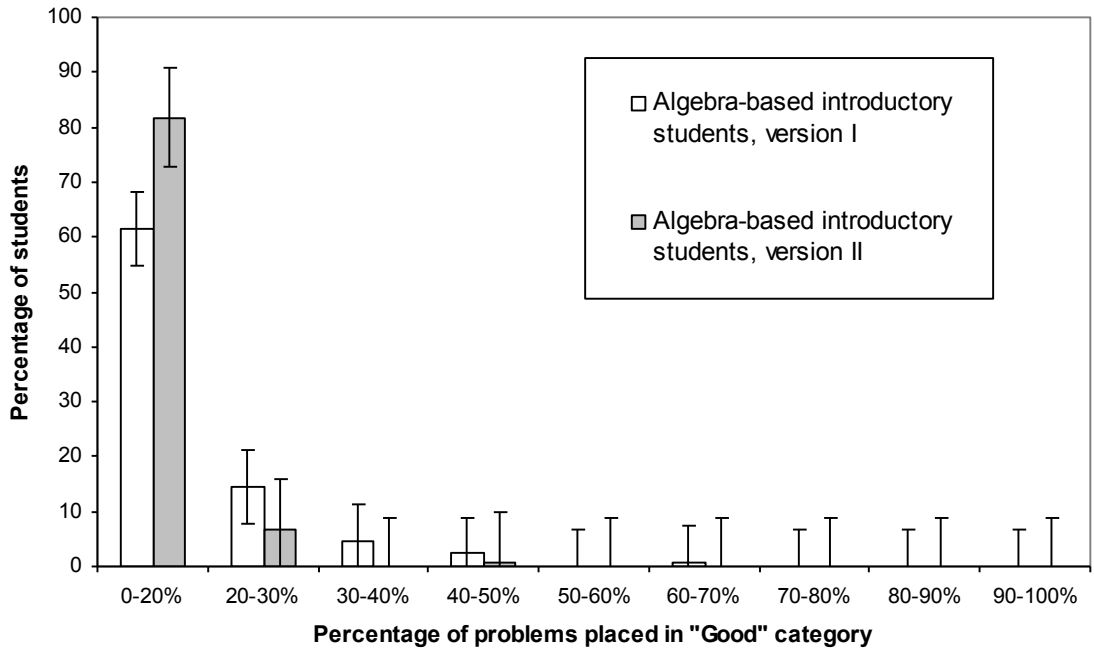

*Figure 2.* Histogram of the algebra-based introductory physics students (for both versions I and II of the problem set) who categorized various percentages of the 25 problems in "good" categories when asked to categorize them based on similarity of solution. Version II involving Chi problems was categorized worse than version I. As discussed in the text, there is no statistically significant difference between the performance of the two algebra-based classes on the 15 problems that were common to the two versions. The error bars refer to the standard error.

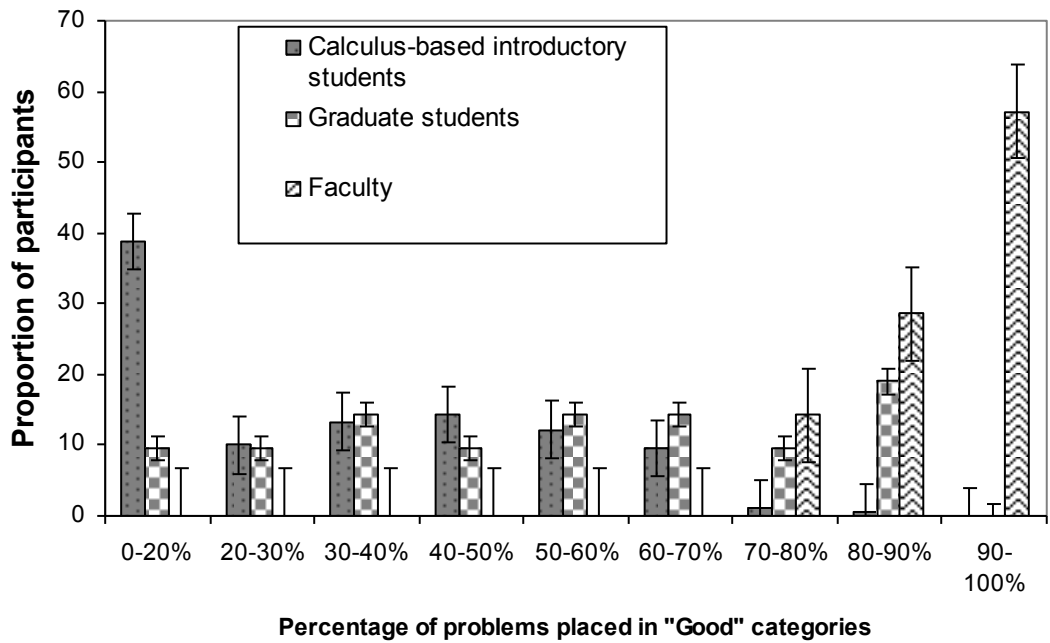

*Figure 3.* **Reprinted with permission from C. Singh, Categorization of problems to assess and improve proficiency as teachers and learners, 77(1), Am. J. Phys., 73 2009.** Histogram of calculus-based introductory physics students, graduate students, and physics faculty who categorized various percentages of the 25 problems in version I in "good" categories when asked to categorize them based on similarity of solution. Physics faculty members performed best in the categorization task followed by graduate students and then introductory physics students, but there is a large overlap between the graduate students and the introductory physics students.

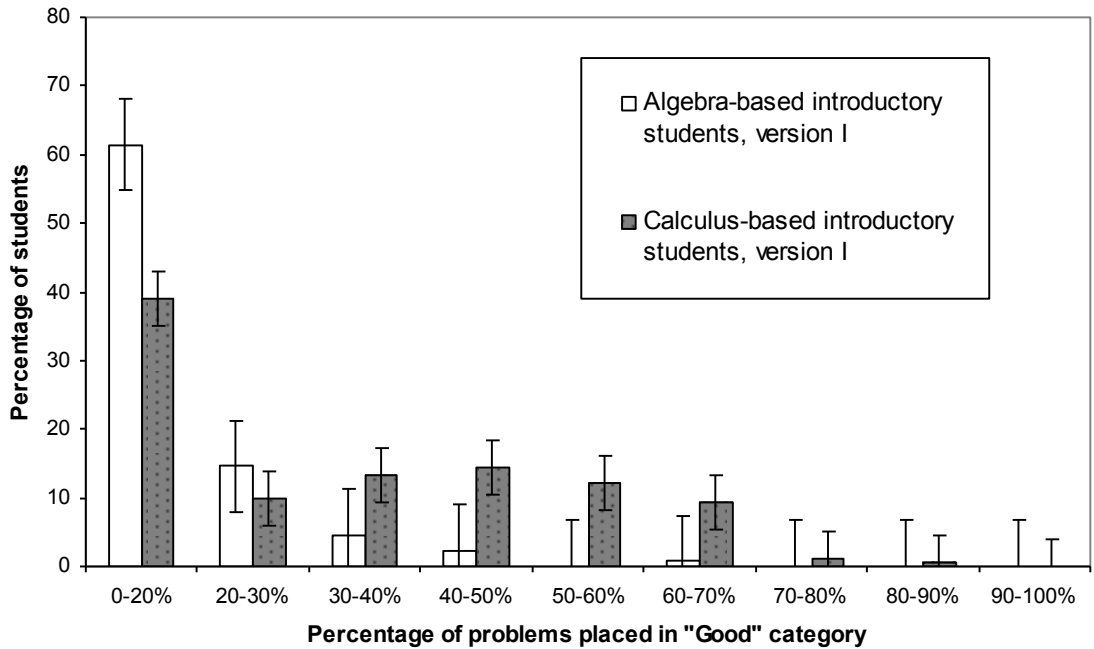

*Figure 4.* Histogram of the calculus-based and algebra-based introductory physics students (version I of the problem set) who categorized various percentages of the 25 problems in "good" categories when asked to categorize them based on similarity of solution. The calculus-based introductory students categorized the problems better than the algebra-based introductory physics students. The error bars refer to the standard error.

# Tables

Table 1 *Examples of Categories for Problem Set Version I*

| Question | Examples of Primary Categories | Examples of Secondary Categories | Poor/Moderate Categories |
|---|---|---|---|
| 1 | (a) momentum conservation or (b) completely inelastic collision | - | speed |
| 2 | (a) mechanical energy conservation or (b) 1D kinematics | - | speed |
| 3 | work by conservative force/definition of work | - | ramp |
| 4 | mechanical energy conservation and momentum conservation | - | energy only or momentum only |
| 5 | mechanical energy conservation and Newton's second law | centripetal acceleration, circular motion/tension | tension only or force only |
| 6 | mechanical energy conservation | - | spring only |
| 7 | work-energy theorem/definition of work or Newton's second law/1D kinematics | relation between kinetic energy and speed | speed |
| 8 | (momentum conservation or completely inelastic collision) and mechanical energy conservation | - | energy only or momentum only |
| 9 | 2D kinematics | - | cliff |
| 10 | Newton's second law | circular motion/friction | friction only |
| 11 | linear momentum conservation or completely inelastic collision | - | speed |
| 12 | mechanical energy conservation and work-energy theorem/definition of work | friction | friction only |
| 13 | Newton's second law | Newton's third law | force |
| 14 | 2D kinematics | - | force/cliff |
| 15 | mechanical energy conservation | - | speed |
| 16 | mechanical energy conservation or 2D kinematics | - | speed |
| 17 | Newton's second law | Newton's third law/tension | tension only |
| 18 | mechanical energy conservation or 2D kinematics | - | speed |
| 19 | impulse-momentum theorem | - | force |
| 20 | mechanical energy conservation or 2D kinematics | - | speed |
| 21 | impulse-momentum theorem | - | force |
| 22 | 2D kinematics | - | ramp |
| 23 | Newton's second law/1D kinematics or Work-energy theorem/definition of work | kinematic variables | force |
| 24 | mechanical energy conservation or momentum conservation or completely inelastic collision | - | speed |
| 25 | mechanical energy conservation and Newton's second law | centripetal acceleration, circular motion/normal force | ramp or force only |

*Note.* These are examples of the primary and secondary categories and one commonly occurring poor/moderate category for each of the 25 questions on the problem set.

Table 2 *Examples of Categories for Problem Set Version II*

| Question | Examples of Primary Categories | Examples of Secondary Categories | Poor/Moderate Categories |
|---|---|---|---|
| 1 (21) [a] | Impulse-momentum theorem | - | Force |
| 2 | Angular momentum conservation | - | Angular speed, moment of inertia |
| 3 (8) | (Momentum conservation or completely inelastic collision) and mechanical energy conservation | - | Energy only or momentum only |
| 4 (13) | Newton's second law | Newton's third law | Force |
| 5 (14) | 2D kinematics | - | Force/cliff |
| 6 (15) | Mechanical energy conservation | - | Speed |
| 7 (17) | Newton's second law | Newton's third law/tension | Tension only |
| 8 (19) | Impulse-momentum theorem | - | Force |
| 9 (24) | Mechanical energy conservation and momentum conservation or completely inelastic collision | - | Speed |
| 10 [b] | Rotational kinematics | Rotational dynamics (implicit) | Angular speed, friction only |
| 11 [b] | Angular momentum conservation | - | Angular speed |
| 12 (22) | 2D kinematics | - | Ramp |
| 13 (12) | Mechanical energy conservation and work-energy theorem/definition of work | Friction | Friction only |
| 14 [b] | (A) Mechanical energy conservation or (B) Newton's second law and kinematics/work energy theorem | - | Speed |
| 15 [b] | Newton's second law | - | Tension only |
| 16 [b] | (A) Mechanical energy conservation /work-energy theorem/definition of work or (B) Newton's second law and kinematics | Friction, potential energy stored in spring, spring force | Friction only, spring only |
| 17 [b] | (A) Work-energy theorem/definition of work or (B) Newton's second law and kinematics | Friction, kinetic energy, gravitational potential energy | Friction only, ramp |
| 18 [b] | (A) Work-energy theorem/definition of work or (B) Newton's second law and kinematics | Friction, kinetic energy, gravitational potential energy | Speed, friction only, ramp |
| 19 | Angular momentum conservation | - | Angular speed, moment of inertia |
| 20 (2) | (A) Mechanical energy conservation or (B) 1D kinematics | - | Speed |
| 21 | Angular momentum conservation | - | Angular speed |
| 22 (4) | Mechanical energy conservation and momentum conservation | - | Energy only or momentum only |
| 23 (5) | Mechanical energy conservation and Newton's second law | Centripetal acceleration, circular motion/tension | Tension only or force only |
| 24 (10) | Newton's second law | Circular motion/friction | Friction only |
| 25 (25) | Mechanical energy conservation and Newton's second law | Centripetal acceleration, circular motion/normal force | Ramp or force only |

*Note.* These are examples of the primary and secondary categories and one commonly occurring poor/moderate category for each of the 25 questions for version II of the problem set. This set includes 7 problems from the Chi study. [a]Refers to a problem which is present in both versions of the problem set. Numbers in parentheses for these problems refer to the problem's number in version I. [b]Refers to a problem from the Chi study.

Table 3 *Performance in Our Study vs. Performance in the Chi Study*

| Chi's Categories | % of 1981 novices (8 total) | % of 1981 experts (8 total) | % of algebra-based students version II (109 total) | | % of algebra-based students version I (114 total) | % of calculus-based students (180 total) |
|---|---|---|---|---|---|---|
| | | | All questions (25) | Chi questions (7) | | |
| Novice Categories from the Chi Study | | | | | | |
| Angular motion (including circular) | 87.5 | - | 72 | 59 | 57 | 42 |
| Inclined planes | 50 | - | 24 | 19 | 19 | 18 |
| Velocity and acceleration | 25 | - | 31 | 26 | 51 | 10.5 |
| Friction | 25 | - | 55 | 51 | 52 | 27 |
| Kinetic energy | 50 | - | 16 | 15 | 15 | 6 |
| Cannot classify/omitted | 50 | - | 44 | 18 | 34 | 39 |
| Vertical motion | 25 | - | 3 | 3 | 3 | 1 |
| Pulleys | 37.5 | - | 16 | 16 | 6 | 2 |
| Free fall | 25 | - | 6 | 1 | 4 | 6 |
| Expert Categories from the Chi Study | | | | | | |
| Newton's 2nd Law (also Newton's Laws) | - | 75 | 22 | 18 | 19 | 38 |
| Energy principles (conservation of energy, work-energy theorem, energy considerations) | - | 75 | 42 | 31 | 35 | 73 |
| Angular motion (not including circular) | - | 75 | 43 | 31 | 39 | 15 |
| Circular motion | - | 62.5 | 29 | 28 | 18 | 27 |
| Statics | - | 50 | 0 | 0 | 0 | 0 |
| Conservation of Angular Momentum | - | 25 | 7 | 1 | 1 | 1 |
| Linear kinematics/motion (not including projectile motion) | - | 25 | 51 | 44 | 42 | 63 |
| Vectors | - | 25 | 1 | 1 | 16 | 2 |
| Categories made by both Novices and Experts from the Chi Study | | | | | | |
| Momentum principles (conservation of momentum, momentum considerations) | 25 | 75 | 39 | 11 | 33 | 64 |
| Work | 50 | 25 | 4 | 4 | 41 | 47 |
| Center of mass | 62.5 | 62.5 | 2 | 0 | 1 | 0 |
| Springs | 75 | 25 | 23 | 23 | 52 | 30 |

*Note.* The novice and expert categories are those made by introductory physics students and graduate students respectively in the Chi study. Introductory physics students in the calculus-based courses (last column) were much more likely than those in the algebra-based courses to place problems in expert-like categories such as Newton's second law, Energy principles, Linear kinematics, Momentum principle and Work. Categories in gray are those for which the questions from the Chi study were not available and our questions

seldom belonged to those categories.